\documentclass[a4paper,reqno,12pt]{amsart}
\usepackage[all]{xy}           
\usepackage{amssymb}           
\usepackage{hyperref}
\usepackage{eucal}
\numberwithin{equation}{section}

\newtheorem{definition}{Definition}[section]
\newtheorem{lemma}[definition]{Lemma}
\newtheorem{theorem}[definition]{Theorem}

\newtheorem{remarkth}[definition]{Remark}

\newenvironment{remark}{\begin{remarkth}\upshape}{\hfill$\diamond$\end{remarkth}}

\renewcommand{\emph}[1]{{\bfseries\itshape{#1}}}

\newcommand{\R}{\mathbb{R}}      

\makeatletter
\newcommand\prol{\@ifstar{\@proldf}{\@prolpf}}  
\def\@prolpf{\@ifnextchar[{\@prolpf@wrt}{\@prolpf@}}
\def\@prolpf@wrt[#1]#2{\@ifnextchar[{\@prolpf@wrt@at{#1}{#2}}{\@prolpf@wrt@{#1}{#2}}}
\def\@prolpf@wrt@at#1#2[#3]{\prolsymbol^{#1}_{#3}#2}
\def\@prolpf@wrt@#1#2{\prolsymbol^{#1}#2}
\def\@prolpf@#1{\@ifnextchar[{\@prolpf@at{#1}}{\@prolpf@@{#1}}}
\def\@prolpf@at#1[#2]{\prolsymbol_{#2}#1}
\def\@prolpf@@#1{\prolsymbol#1}
\def\@proldf{\@ifnextchar[{\@proldf@wrt}{\@proldf@}}
\def\@proldf@wrt[#1]#2{\@ifnextchar[{\@proldf@wrt@at{#1}{#2}}{\@proldf@wrt@{#1}{#2}}}
\def\@proldf@wrt@at#1#2[#3]{\prolsymbol^{*#1}_{#3}#2}
\def\@proldf@wrt@#1#2{\prolsymbol^{*#1}#2}
\def\@proldf@#1{\@ifnextchar[{\@proldf@at{#1}}{\@proldf@@{#1}}}
\def\@proldf@at#1[#2]{\prolsymbol^*_{#2}#1}
\def\@proldf@@#1{\prolsymbol^*#1}
\def\prolsymbol{\mathcal{T}}
\makeatother

\begin{document}

\title[HAMILTON-JACOBI THEORY FOR CLASSICAL FIELD THEORIES]
{A GEOMETRIC HAMILTON-JACOBI THEORY FOR CLASSICAL FIELD THEORIES}

\author[M. de Le\'on]{Manuel de Le\'on}
\address{Instituto de Ciencias Matem\'aticas, CSIC-UAM-UC3M-UCM,
Serrano 123, 28006
Madrid, Spain} \email{mdeleon@imaff.cfmac.csic.es}

\author[J.C.\ Marrero]{Juan Carlos Marrero}
\address{J.C.\ Marrero:
Departamento de Matem\'atica Fundamental, Universidad de La Laguna,
La Laguna, Canary Islands, Spain} \email{jcmarrer@ull.es}

\author[D.\ Mart\'{\i}n de Diego]{David Mart\'{\i}n de Diego}
\address{D.\ Mart\'{\i}n de Diego:
Instituto de Ciencias Matem\'aticas, CSIC-UAM-UC3M-UCM, Serrano 123,
28006 Madrid, Spain} \email{d.martin@imaff.cfmac.csic.es}

\keywords{Multisymplectic field theory, Hamilton-Jacobi equations}\footnote{\bf To Prof. Demeter Krupka
in his 65th birthday}

 \subjclass[2000]{70S05, 49L99}

\begin{abstract}
In this paper we extend the geometric formalism of the
Hamilton-Jacobi theory for hamiltonian mechanics to the case of
classical field theories in the framework of multisymplectic
geometry and Ehresmann connections.
\end{abstract}

\thanks{This work has been partially supported by MEC (Spain) Grants MTM
2006-03322, MTM 2007-62478, project ``Ingenio Mathematica" (i-MATH)
No. CSD 2006-00032 (Consolider-Ingenio 2010) and S-0505/ESP/0158 of
the CAM}

 \maketitle

\tableofcontents

\section{Introduction}

The standard formulation of the Hamilton-Jacobi problem is to find a
function $S(t, q^A)$ (called the {\bf principal function}) such that
\begin{equation}\label{hj1}
\frac{\partial S}{\partial t} + H(q^A, \frac{\partial S}{\partial
q^A}) = 0.
\end{equation}
If we put $S(t, q^A) = W(q^A) - t E$, where $E$ is a constant, then
$W$ satisfies
\begin{equation}\label{hj2}
H(q^A, \frac{\partial W}{\partial q^A}) = E;
\end{equation}
$W$ is called the {\bf characteristic function}.

Equations (\ref{hj1}) and (\ref{hj2}) are indistinctly referred as
the {\bf Hamilton-Jacobi equation}.

There are some recent attempts to extend this theory for classical
field theories in the framework of the so-called multisymplectic
formalism \cite{paufler1,paufler2}. For a classical field theory the
hamiltonian is a function $H=H(x^\mu, y^i, p_i^\mu)$, where
$(x^\mu)$ are coordinates in the space-time, $(y^i)$ represent the
field coordinates, and $(p_i^\mu)$ are the conjugate momenta.

In this context, the Hamilton-Jacobi equation is \cite{rund}
\begin{equation}\label{hj3}
\frac{\partial S^\mu}{\partial x^\mu} + H(x^\nu, y^i, \frac{\partial
S^\mu}{\partial y^i})  = 0
\end{equation}
where $S^\mu = S^\mu(x^\nu, y^j)$.

In this paper we introduce a geometric version for the
Hamilton-Jacobi theory based in two facts: (1) the recent
geometric description for Hamiltonian mechanics developed in
\cite{carinena} (see \cite{hjnoholonomic} for the case of
nonholonomic mechanics); (2) the multisymplectic formalism for
classical field theories \cite{BSF,CIL1,CIL2,gimmsy} in terms of
Ehresmann connections \cite{LeMaMa1,LeMaMa2,campos1,campos2}.

We shall also adopt the convention that  a repeated index implies
summation over the range of the index.

\section{A geometric Hamilton-Jacobi theory for Hamiltonian
mechanics}

First of all, we give a geometric version of the standard
Hamilton-Jacobi theory which will be useful in the sequel.

Let $Q$ be the configuration manifold, and $T^*Q$ its cotangent
bundle equipped with the canonical symplectic form
$$
\omega_Q = dq^A \wedge dp_A
$$
where $(q^A)$ are coordinates in $Q$ and $(q^A, p_A)$ are the
induced ones in $T^*Q$.

Let $H : T^*Q \longrightarrow \R$ a hamiltonian function and $X_H$
the corresponding hamiltonian vector field:
$$
i_{X_H} \, \omega_Q = dH
$$

The integral curves of $X_H$, $(q^A(t), p_A(t))$, satisfy the
Hamilton equations:
$$
\frac{dq^A}{dt} = \frac{\partial H}{\partial p_A} \, , \; \frac{dp_
A}{dt} = - \frac{\partial H}{\partial q^A}
$$

\begin{theorem}[Hamilton-Jacobi Theorem]
Let $\lambda$ be a closed 1-form on $Q$ (that is, $d\lambda=0$ and,
locally $\lambda = dW$). Then, the following conditions are
equivalent:

\begin{enumerate}
\item[(i)] If $\sigma: I\to Q$ satisfies the equation
$$
\frac{dq^A}{dt} = \frac{\partial H}{\partial p_A}
$$
then $\lambda\circ \sigma$ is a solution of the Hamilton equations;

\item[(ii)] $d (H\circ \lambda)=0$.
\end{enumerate}
\end{theorem}

To go further in this analysis, define a vector field on $Q$:
$$
X_H^{\lambda}=T\pi_Q\circ X_H\circ \lambda
$$
as we can see in the following diagram:

\[ \xymatrix{ T^*Q
\ar[dd]^{\pi_Q} \ar[rrr]^{X_H}&   & &T(T^*Q)\ar[dd]^{T\pi_Q}\\
  &  & &\\
 Q\ar@/^2pc/[uu]^{\lambda}\ar[rrr]^{X_H^{\lambda}}&  & & TQ }
\]

Notice that the following conditions are equivalent:
\begin{enumerate}
\item[(i)] If $\sigma: I\to Q$ satisfies the equation
$$
\frac{dq^A}{dt} = \frac{\partial H}{\partial p_A}
$$
then $\lambda\circ \sigma$ is a solution of the Hamilton equations;
\item[(i)'] If $\sigma: I\to Q$ is an integral curve of
$X_H^{\lambda}$, then $\lambda\circ \sigma$ is an integral curve of
$X_H$;
\item[(i)''] $X_H$ and $X_H^{\lambda}$ are $\lambda$-related,
i.e.
$$
T\lambda(X_H^{\lambda})=X_H \circ \lambda
$$
\end{enumerate}
so that the above theorem can be stated as follows:

\begin{theorem}[Hamilton-Jacobi Theorem]
Let $\lambda$ be a closed 1-form on $Q$. Then, the following
conditions are equivalent:
\begin{enumerate}
\item[(i)] $X_H^{\lambda}$ and $X_H$ are $\lambda$-related;

\item[(ii)] $d (H\circ \lambda)=0$.
\end{enumerate}
\end{theorem}

\section{The multisymplectic formalism}

\subsection{Multisymplectic bundles}

The configuration manifold in Mechanics is substituted by a fibred
manifold
$$
\pi : E \longrightarrow M
$$
such that
\begin{enumerate}
\item $\dim M = n, \; \dim E = n+m$ \item $M$ is endowed with a
volume form $\eta$.
\end{enumerate}

We can choose fibred coordinates $(x^\mu, y^i)$ such that
$$
\eta = dx^1 \wedge \cdots \wedge dx^n\; .
$$

We will use the following useful notations:
\begin{eqnarray*}
&&d^nx = dx^1 \wedge \cdots \wedge dx^n\\
&& d^{n-1}x^\mu = i_{\frac{\partial}{\partial x^\mu}} \, d^nx\; .
\end{eqnarray*}

Denote by $ V\pi = \ker T\pi$ the vertical bundle of $\pi$, that is,
their elements are the tangent vectors to $E$ which are
$\pi$-vertical.

Denote by
$$
\Pi: \Lambda^n E \longrightarrow E
$$
the vector bundle of $n$-forms on $E$.

The total space $\Lambda^nE$ is equipped with a canonical $n$-form
$\Theta$:
$$
\Theta(\alpha)(X_1, \dots, X_n) = \alpha(e)(T\Pi(X_1), \dots, T\Pi
(X_n))
$$
where $X_1, \dots, X_n \in T_{\alpha}(\Lambda^nE)$ and $\alpha$ is
an $n$-form at $e \in E$.

The $(n+1)$-form
$$
\Omega = - d \Theta\; ,
$$
is called the canonical multisymplectic form on $\Lambda^nE$.

Denote by $\Lambda^n_r E$ the bundle of $r$-semibasic $n$-forms on
$E$, say
$$
\Lambda^n_r E = \{\alpha \in \Lambda^n E \; | \; i_{v_1 \wedge
\cdots \wedge v_r} \alpha = 0, \; \hbox{whenever $v_1, \dots, v_r$
are $\pi$-vertical}  \}
$$
Since $\Lambda^n_r E$ is a submanifold of $\Lambda^n E$ it is
equipped with a multisymplectic form $\Omega_r$, which is just
the restriction of $\Omega$.

Two bundles of semibasic forms play an special role: $\Lambda ^n_1
E$ and $\Lambda^n_2E$. The elements of these spaces have the
following local expressions:
\begin{eqnarray*}
\Lambda^n_1 E \; &:& \; p_0 \, d^nx \\
\Lambda^n_2 E \; &:& p_0 \, d^nx + p^\mu_i \, dy^i \wedge
d^{n-1}x^\mu\; .
\end{eqnarray*}
which permits to introduce local coordinates $(x^\mu, y^i, p_0)$ and
$(x^\mu, y^i, p_0, p^\mu_i)$ in $\Lambda^n_1 E$ and $\Lambda^n_2 E$,
respectively.

Since $\Lambda^n_1E$ is a vector subbundle of $\Lambda^n_2E$ over
$E$, we can obtain the quotient vector space denoted by $J^1\pi^*$
which completes the following exact sequence of vector bundles:
$$
0 \longrightarrow \Lambda^n_1E \longrightarrow \Lambda^n_2E
\longrightarrow J^1 \pi^* \longrightarrow 0\; .
$$
We denote by  $\pi_{1,0}: J^1\pi^* \longrightarrow E$ and
$\pi_{1}: J^1\pi^* \longrightarrow M$ the induced fibrations.

\subsection{Ehresmann Connections in the fibration $\pi_{1}: J^1\pi^* \longrightarrow
M$}

A \emph{connection} (in the sense of Ehresmann) in $\pi_1$ is a
horizontal subbundle ${\bf H}$ which is complementary to $V\pi_1$;
namely,
$$
T(J^1\pi^*)= {\bf H} \oplus V\pi_1
$$
where $V\pi_1 = \ker T\pi_1$ is the vertical bundle of $\pi_1$.
Thus, we have:
\begin{enumerate}
\item there exists a (unique) horizontal lift of every tangent
vector to $M$;
\item in  fibred coordinates $(x^\mu, y^i, p^\mu_i)$ on $J^1\pi^*$, then
$$
V\pi_1 =\hbox{span } \{\frac{\partial}{\partial y^i},
\frac{\partial}{\partial p^\mu_i}\} \; , {\bf H} =\hbox{span }
\{{\bf H}_\mu \}\; ,
$$
where ${\bf H}_\mu$ is the horizontal lift of
$\frac{\partial}{\partial x^\mu}$.
\item there is a horizontal projector ${\bf h} : TJ^* \pi
\longrightarrow {\bf H}$.
\end{enumerate}

\subsection{Hamiltonian sections}

Consider a hamiltonian section
$$
h : J^1\pi^* \longrightarrow \Lambda^n_2E
$$
of the canonical projection $\mu: \Lambda^n_2E \longrightarrow
J^1\pi^*$ which in local coordinates read as
$$
h(x^\mu, y^i, p^\mu_i) = (x^\mu, y^i, -H(x, y, p), p^\mu_i)\; .
$$

Denote by $\Omega_h = h^* \Omega_{2}$, where $\Omega_2$ is the
multisymplectic form on $\Lambda^n_2E$.

The field equations can be written as follows:
\begin{equation}\label{fieldequation}
i_{\bf h} \, \Omega_h = (n-1) \, \Omega_h\; ,
\end{equation}
where ${\bf h}$ denotes the horizontal projection of an Ehresmann
connection in the fibred manifold $\pi_{1}: J^1\pi^* \longrightarrow
M$.

The local expressions of $\Omega_2$ and $\Omega_h$ are:
\begin{eqnarray*}
 \Omega_2 &=& - d(p_0 \, d^n x + p^\mu_i \,
dy^i \wedge
d^{n-1}x^\mu)\\
\Omega_h &=& - d(- H \, d^n x + p^\mu_i \, dy^i \wedge
d^{n-1}x^\mu)\; .
\end{eqnarray*}

\subsection{The field equations}

Next, we go back to the Equation (\ref{fieldequation}).

The horizontal subspaces are locally spanned by the local vector
fields
$$
H_\mu = {\bf h}(\frac{\partial}{\partial x^\mu}) =
\frac{\partial}{\partial x^\mu} + \Gamma_\mu^i \,
\frac{\partial}{\partial y^i} + (\Gamma_\mu)_j^\nu \,
\frac{\partial}{\partial p^\nu_j},
$$
where $\Gamma_\mu^i$ and $(\Gamma_\mu)_j^\nu$ are the Christoffel
components of the connection.

Assume that $\tau$ is an integral section of ${\bf h}$; this means
that $\tau : M \longrightarrow J^1\pi^*$ is a local section of the
canonical projection $\pi_{1} : J^1\pi^* \longrightarrow M$ such
that $T\tau(x) (T_xM) = {\bf H}_{\tau(x)}$, for all $x \in M$.

If $ \tau(x^\mu) = (x^\mu, \tau^i(x), \tau^\mu_i(x)) $ then the
above conditions becomes
$$
\frac{\partial \tau^i}{\partial x^\mu} = \frac{\partial H}{\partial
p^\mu_i} \; , \; \frac{\partial \tau^\mu_i}{\partial x^\mu} = -
\frac{\partial H}{\partial y^i}
$$
which are the Hamilton equations.

\section{The Hamilton-Jacobi theory}

Let $\lambda$ be a $2$-semibasic $n$-form on $E$; in local
coordinates we have
$$
\lambda = \lambda_0 (x, y) \, d^nx + \lambda^\mu_i(x,y) \, dy^i
\wedge d^{n-1}x^\mu\; .
$$

Alternatively, we can see it as a section $\lambda : E
\longrightarrow \Lambda^n_2 E$, and then we have
$$
\lambda (x^\mu, y^i) = (x^\mu, y^i, \lambda_0(x,y),
\lambda^\mu_i(x,y))\; .
$$

A direct computation shows that
$$
d\lambda = \left(\frac{\partial \lambda_0}{\partial y^i} -
\frac{\partial \lambda^\mu_i}{\partial x^\mu} \right) \, dy^i
\wedge d^nx + \frac{\partial \lambda^\mu_i}{\partial y^j} \, dy^j
\wedge dy^i \wedge d^{n-1}x^\mu\; .
$$
Therefore, $d \lambda = 0$ if and only if
\begin{eqnarray}
\frac{\partial \lambda_0}{\partial y^i} &=& \frac{\partial \lambda^\mu_i}{\partial x^\mu}\\
\frac{\partial \lambda^\mu_i}{\partial y^j} &=& \frac{\partial
\lambda^\mu_j}{\partial y^i}\; .
\end{eqnarray}

Using $\lambda$ and ${\bf h}$ we construct an induced connection in
the fibred manifold $\pi : E \longrightarrow M$ by defining its
horizontal projector as follows:
\begin{eqnarray*}
\tilde{h}_e & : & T_eE \longrightarrow T_eE\\
\tilde{h}_e (X) &=& T\pi_{1,0} \circ h_{(\mu \circ \lambda)(e)}
\circ \epsilon (X)
\end{eqnarray*}
where $\epsilon (X) \in T_{(\mu \circ \lambda)(e)}(J^1\pi^*)$ is
an arbitrary tangent vector which projects onto $X$.

{}From the above definition we immediately proves that
\begin{enumerate}
\item $\tilde{\bf h}$ is a well-defined  connection in the fibration $\pi: E\longrightarrow M$.

\item The corresponding horizontal subspaces are locally spanned by
$$
\tilde{H}_\mu = \tilde{h}(\frac{\partial}{\partial x^\mu}) =
\frac{\partial}{\partial x^\mu} + \Gamma_\mu^i ((\mu \circ
\lambda)(x,y))\, \frac{\partial}{\partial y^i}\; .
$$

\end{enumerate}

The following theorem is the main result of this paper.

\begin{theorem}
Assume that $\lambda$ is a closed 2-semibasic form on $E$ and that
$\tilde{h}$ is a flat connection on $\pi: E \longrightarrow M$.
Then the following conditions are equivalent:

\begin{itemize}
\item[(i)] If $\sigma$ is an integral section of $\tilde{h}$ then
$\mu \circ \lambda \circ \sigma$ is a solution of the Hamilton
equations.
\item[(ii)] The $n$-form $h \circ \mu \circ \lambda$ is closed.
\end{itemize}

\end{theorem}

Before to begin with the proof, let us consider some preliminary
results.

We have
$$
(h \circ \mu \circ \lambda)(x^\mu, y^i) = (x^\mu, y^i, - H(x^\mu,
y^i, \lambda^\mu_i(x,y)), \lambda^\mu_i(x, y))\; ,
$$
that is
$$ h \circ \mu \circ \lambda = - H(x^\mu, y^i,
\lambda^\mu_i(x,y)) \, d^nx + \lambda^\mu_i \, dy^i \wedge
d^{n-1}x^\mu\; .
$$

Notice that $h \circ \mu \circ \lambda$ is again a $2$-semibasic
$n$-form on $E$.

A direct computation shows that
\begin{eqnarray*}
d(h \circ \mu \circ \lambda) &=& - \left(\frac{\partial H}{\partial
y^i} + \frac{\partial H}{\partial p^\nu_j} \frac{\partial
\lambda^\nu_j}{\partial y^i} + \frac{\partial
\lambda^\mu_i}{\partial x^\mu}\right) \, dy^i \wedge d^nx \\
&& + \frac{\partial \lambda^\mu_i}{\partial y^j} \, dy^j \wedge
dy^i \wedge d^{n-1}x^\mu\; .
\end{eqnarray*}

Therefore, we have the following result.

\begin{lemma}\label{lemma}
Assume $d\lambda = 0$; then
$$
d(h \circ \mu \circ \lambda) = 0
$$
if and only if
$$
\frac{\partial H}{\partial y^i} + \frac{\partial H}{\partial
p^\nu_j} \frac{\partial \lambda^\nu_j}{\partial y^i} +
\frac{\partial \lambda^\mu_i}{\partial x^\mu} = 0\; .
$$
\end{lemma}

{\bf Proof of the Theorem}

 $(i) \Rightarrow (ii)$

It should be remarked the meaning of $(i)$.

Assume that
$$
\sigma (x^\mu) = (x^\mu, \sigma^i(x))
$$
is an integral section of $\tilde{\bf h}$; then
$$ \frac{\partial \sigma^i}{\partial x^\mu} = \frac{\partial
H}{\partial p^\mu_i}\; .
$$

$(i)$ states that in the above conditions,
$$
(\mu \circ \lambda \circ \sigma) (x^\mu) = (x^\mu, \sigma^i(x),
\bar{\sigma}^\nu_j = \lambda^\nu_j(\sigma(x)))
$$
is a solution of the Hamilton equations, that is,
$$
\frac{\partial \bar{\sigma}^\mu_i}{\partial x^\mu} =
\frac{\partial \lambda^\mu_i}{\partial x^\mu} + \frac{\partial
\lambda^\mu_i}{\partial y^j}\frac{\partial \sigma^j}{\partial
x^\mu} = - \frac{\partial H}{\partial y^i}.
$$

Assume $(i)$. Then
\begin{eqnarray*}
&& \frac{\partial H}{\partial y^i} + \frac{\partial H}{\partial
p^\nu_j} \frac{\partial \lambda^\nu_j}{\partial y^i} +
\frac{\partial \lambda^\mu_i}{\partial x^\mu} \\
&& = \frac{\partial H}{\partial y^i} + \frac{\partial H}{\partial
p^\nu_j} \frac{\partial \lambda^\nu_i}{\partial y^j} +
\frac{\partial \lambda^\mu_i}{\partial x^\mu}\, , \qquad (\hbox{since $d\lambda = 0$})\\
&& = \frac{\partial H}{\partial y^i} + \frac{\partial
\sigma^j}{\partial x^\nu} \frac{\partial \lambda^\nu_i}{\partial
y^j} + \frac{\partial \lambda^\mu_i}{\partial x^\mu}\, ,  \qquad (\hbox{since the first Hamilton equation})\\
&& = 0 \qquad (\hbox{since $(i)$})
\end{eqnarray*}
which implies $(ii)$ by Lemma \ref{lemma}.

$(ii) \Rightarrow (i)$

Assume that $d(h \circ \mu \circ \lambda) = 0$.

Since $\tilde{h}$ is a flat connection, we may consider an integral
section $\sigma$ of $\tilde{h}$. Suppose that
$$
\sigma(x^\mu) = (x^\mu, \sigma^i(x)). $$ Then, we have that
$$
\frac{\partial \sigma^i}{\partial x^\mu} = \frac{\partial
H}{\partial p^\mu_i}.
$$
Thus,
\begin{eqnarray*}
\frac{\partial \bar{\sigma}^\mu_j}{\partial x^\mu} &=&
\frac{\partial \lambda^\mu_j}{\partial x^\mu} + \frac{\partial
\lambda^\mu_j}{\partial y^i} \frac{\partial \sigma^i}{\partial
x^\mu}\, ,\\
&=&  \frac{\partial \lambda^\mu_j}{\partial x^\mu} +
\frac{\partial \lambda^\mu_i}{\partial y^j} \frac{\partial
\sigma^i}{\partial
x^\mu}\, , \qquad  (\hbox{since $d\lambda = 0$})\\
&=& \frac{\partial \lambda^\mu_j}{\partial x^\mu} + \frac{\partial
\lambda^\mu_i}{\partial y^j} \frac{\partial H}{\partial
p^\mu_i}\, ,  \qquad (\hbox{since the first Hamilton equation})\\
&& = - \frac{\partial H}{\partial y^j}\, ,  \qquad (\hbox{since
$(ii)$}).\qquad \hfill \Box
\end{eqnarray*}

\bigskip

Assume that $\lambda = dS$, where $S$ is a $1$-semibasic
$(n-1)$-form, say
$$
S = S^\mu \, d^{n-1} x^\mu
$$
Therefore, we have
$$
\lambda_0 = \frac{\partial S^\mu}{\partial x^\mu} \; , \;
\lambda^\mu_i = \frac{\partial S^\mu}{\partial y^i}
$$
and the Hamilton-Jacobi equation has the form
$$
\frac{\partial}{\partial y^i} \left( \frac{\partial
S^\mu}{\partial x^\mu} + H(x^\nu, y^i, \frac{\partial
S^\mu}{\partial y^i}) \right) = 0\; .
$$

The above equations mean that
$$
\frac{\partial S^\mu}{\partial x^\mu} + H(x^\nu, y^i, \frac{\partial
S^\mu}{\partial y^i}) = f(x^\mu)
$$
so that if we put $\tilde{H} = H - f$ we deduce the standard form of
the Hamilton-Jacobi equation (since $H$ and $\tilde{H}$ give the
same Hamilton equations):
$$
\frac{\partial S^\mu}{\partial x^\mu} + \tilde{H}(x^\nu, y^i,
\frac{\partial S^\mu}{\partial y^i})  = 0\; .
$$

An alternative geometric approach of the Hamilton-Jacobi theory
for Classical Field Theories in a multisymplectic setting was
discussed in \cite{paufler1,paufler2}.

\section{Time-dependent mechanics}

A hamiltonian time-dependent mechanical system corresponds to a
classical field theory when the base is $M = \R$.

We have the following identification $ \Lambda^1_2 E = T^*E $ and
we have local coordinates $(t,y^i, p_0, p_i)$ and $(t, y^i, p_i)$
on $T^*E$ and $J^1\pi^*$, respectively. The hamiltonian section is
given by
$$
h(t, y^i, p_i) = (t, y^i, - H(t,y,p), p_i)\; ,
$$
and therefore we obtain
$$
\Omega_h = dH \wedge dt - dp_i \wedge dy^i\; .
$$

If we denote by $\eta=dt$ the different pull-backs of $dt$ to the
fibred manifolds over $M$, we have the following result.

The pair $(\Omega_h, dt)$ is a cosymplectic structure on $E$, that
is, $\Omega_{h}$ and $dt$ are closed forms and $dt \wedge
\Omega_{h}^n = dt\wedge \Omega_{h}\wedge \dots \wedge \Omega_{h}$
is a volume form, where $dim E = 2n+1$. The Reeb vector field
${\mathcal R}_h$ of the structure $(\Omega_{h}, dt)$ satisfies
$$
i_{{\mathcal R}_h} \, \Omega_h = 0 \; , \, i_{{\mathcal R}_h} \, dt = 1.
$$
The integral curves of ${\mathcal R}_h$ are just the solutions of
the Hamilton equations for $H$.

The relation with the multisymplectic approach is the following:
$$
{\bf h} = {\mathcal R}_h \otimes dt\; ,
$$
or, equivalently,
$$
{\bf h}(\frac{\partial}{\partial t}) = {\mathcal R}_h\; .
$$

A closed $1$-form $\lambda$ on $E$ is locally represented by
$$
\lambda = \lambda_0 \, dt + \lambda_i \, dy^i.
$$
 Using $\lambda$ we obtain a vector field on $E$:
$$
({\mathcal R}_h)_\lambda = T \pi_{1,0} \circ {\mathcal R}_h \circ \mu \circ
\lambda
$$
such that the induced connection is
$$
\tilde{\bf h} = ({\mathcal R}_h)_\lambda \otimes dt
$$

Therefore, we have the following result.

\begin{theorem}\label{H-Jtimedependent}
The following conditions are equivalent:

\begin{itemize}
\item[(i)] $({\mathcal R}_h)_\lambda$ and ${\mathcal R}_h$ are $(\mu \circ
\lambda)$-related.
\item[(ii)] The $1$-form $h \circ \mu \circ \lambda$ is closed.
\end{itemize}
\end{theorem}

\begin{remark}{\rm
An equivalent result to Theorem \ref{H-Jtimedependent} was proved
in \cite{MaSo} (see Corollary 5 in \cite{MaSo}). }
\end{remark}

Now, if
$$
\lambda = dS = \frac{\partial S}{\partial t} \, dt +
\frac{\partial S}{\partial y^i} \, dy^i\; ,
$$
then we obtain the Hamilton-Jacobi equation
$$
\frac{\partial}{\partial y^i} \left(\frac{\partial S}{\partial t}
+ H(t, y^i, \frac{\partial S}{\partial y^i}) \right) = 0\; .
$$

\end{document}